# Towards a Governance Framework for Brain Data

Authors:

Marcello Ienca[1,18]*, Joseph J. Fins[2], Ralf J. Jox[3], Fabrice Jotterand[4,17], Silja Voeneky[5], Roberto Andorno[6], Tonio Ball[7], Claude Castelluccia[8], Ricardo Chavarriaga[9], Hervé Chneiweiss[10], Agata Ferretti[1], Orsolya Friedrich[11], Samia Hurst[12], Grischa Merkel[13], Fruzsina Molnár-Gábor[14], Jean-Marc Rickli[15], James Scheibner[1], Effy Vayena[1], Rafael Yuste[16]*, Philipp Kellmeyer[7]*

**Abstract**

The increasing availability of brain data within and outside the biomedical field, combined with the application of artificial intelligence (AI) to brain data analysis, poses a challenge for ethics and governance. We identify distinctive ethical implications of brain data acquisition and processing, and outline a multi-level governance framework. This framework is aimed at maximizing the benefits of facilitated brain data collection and further processing for science and medicine whilst minimizing risks and preventing harmful use. The framework consists of four primary areas of regulatory intervention: binding regulation, ethics and soft law, responsible innovation, and human rights.

[1] Department of Health Sciences and Technologies, ETH Zurich, Switzerland
[2] New York Presbyterian Hospital and Weill Cornell Medical College, USA
[3] Institute of Humanities in Medicine, Lausanne University Hospital, Switzerland
[4] Medical College of Wisconsin, USA
[5] University of Freiburg, Law Faculty, Dep. International Law and Ethics of Law, Germany
[6] Faculty of Law and Institute for Biomedical Ethics, University of Zurich, Switzerland
[7] Department for Neurosurgery, University Medical Center Freiburg, Germany
[8] Privatics Group, INRIA, France
[9] ZHAW School of Engineering, Switzerland; IEEE Standards Association; Confederation of Laboratories for AI Research in Europe (CLAIRE)
[10] Centre de Recherche Neuroscience Paris Seine, CNRS, France; UNESCO Chair of Bioethics
[11] Institute for Philosophy, FernUniversität in Hagen, Germany
[12] Institute for Ethics, History, and the Humanities, University of Geneva, Switzerland
[13] Law Faculty, University of Kiel, Germany
[14] Heidelberg Academy of Sciences and Humanities, Germany
[15] Geneva Center for Security Policy, Switzerland
[16] The NeuroTechnology Center, Columbia University, USA
[17] Institute of Biomedical Ethics, University of Basel, Switzerland
[18] College of Humanities, EPFL, Switzerland
* Corresponding authors

**Text**

Human brain data are becoming a sought-after commodity in an increasing number of contexts and activities. Until a few years ago their acquisition and analysis were limited to the clinical field and biomedical, psychological or behavioral research. Today, brain data are also increasingly being used in employment, education, and military contexts, as well as for personal use through an increasing number of consumer-grade devices.

In the consumer space, information technology companies are developing devices and applications that leverage brain data for consumer purposes such as cognitive monitoring, neurofeedback, device control or other forms of brain-computer interfacing. For example, in 2017 Facebook launched a brain-computer interface (BCI) research program that aims to build a wearable BCI that enables users to type by simply imagining speech. Microsoft is working in parallel on non-invasive interactive BCIs for the general population while a whole ecosystem of neurotechnology companies such as *Neuralink, Emotiv* and *Kernel* is rapidly emerging. Consumer neurotechnology, e-learning, digital phenotyping, affective computing, psychographics and neuromarketing are some of the domains of application that leverage brain data as a commodity [1, 2].

In the educational and work setting, attempts have been made to collect and process brain data for purposes such as improving learning and redesigning workflows. For example, last year, in China, primary school children were enrolled in a trial where electroencephalography (EEG) data were recorded during cognitive tasks to assess their attention spans[3]. Also in China, government-backed workplace surveillance projects are deploying personal neurotechnologies to detect changes in brain activity among factory employees on the production line. These neurotechnologies are intended to monitor productivity and adjust the pace of production accordingly [4].

Finally, military uses of neurotechnologies and the associated acquisition of brain data have increased in quantity and variety. One example is the "Next-generation Nonsurgical Neurotechnology Program" (N³), a $104 million effort launched in 2019 by the United States Defense Advanced Research Projects Agency (DARPA) with the aim of developing non-invasive, portable and bidirectional BCIs for able-bodied service members [5]. Several other nations have military research programs that involve brain data [6].

These novel uses of brain data add to the already extensive use of these data in clinical medicine and biomedical research. In these fields, electrophysiology and neuroimaging datasets have steadily grown in volume, variety and analytic complexity [7, 8]. Data repurposing, a frequent occurrence in digital health and digital phenotyping, also permits cross-domain data transfer, blurring the lines between biomedical and non-medical data uses.

### An Ethical and Policy Challenge

The increasing availability of brain data inside and outside the biomedical and health-care domain raises challenges for regulation and governance. On the one hand, expanding the volume and variety of brain data available for research is crucial for advancing our scientific understanding of the human brain and providing preventive, diagnostic and therapeutic solutions for patients with neurological or psychiatric disorders [9, 10]. Several large-scale research programs, such as the US BRAIN Initiative and the EU Human Brain Project, are working on advancing measurement tools and computational methods in neuroscience and neurotechnology. These projects could benefit from increased data availability in the medical or consumer domain.

On the other hand, as brain data become part of a wider digital ecosystem, they are subject to the same risks and vulnerabilities as other digital data. These include re-identification, hacking, unauthorized reuse, asymmetric commodification, privacy-sensitive data mining, digital surveillance and co-opting data for other non-benign purposes [2, 11]. Most importantly, brain-related measurements in the non-medical domain are rarely available in isolation. They can be combined with other digitally available information and contextualized against online queries, social media, self-tracked data, DNA and geolocation. Advances in big data analytics and machine learning (ML) portend an unprecedented capacity to infer and identify patterns and predict outcomes by aggregating data from multiple sources [12-16].

Given the increased availability of brain data and recent emphasis in national and international policymaking on data governance, the following question arises: how should brain data be regulated? In particular: what kind of governance framework is needed to maximize the benefits of brain data processing for scientific research and medicine whilst ensuring ethical use in other areas?

**What makes brain data important?**

The notion of "brain data" is often used without a clear conceptual characterization. To promote clarity for regulation and governance purposes, we propose the following working definition: *Human brain data* are *quantitative data about human brain structure, activity and function.* These include direct measurements of brain structure, activity and/or function (e.g., neuronal firing or summed bioelectric signals from EEG) and indirect functional indicators (i.e., blood flow in fMRI and fNIRS). These types of brain data can be combined with non-neural contextual data, such as voice recordings or smartphone usage data, that can be used to support inferences about mental processes in a broader sense (Fig 1). Compared to other measurements of the human body, the risks associated with the collection and processing of brain data are distinctive in terms of quality and magnitude. This is due to inherent properties of brain data and their resulting ethical and legal implications.

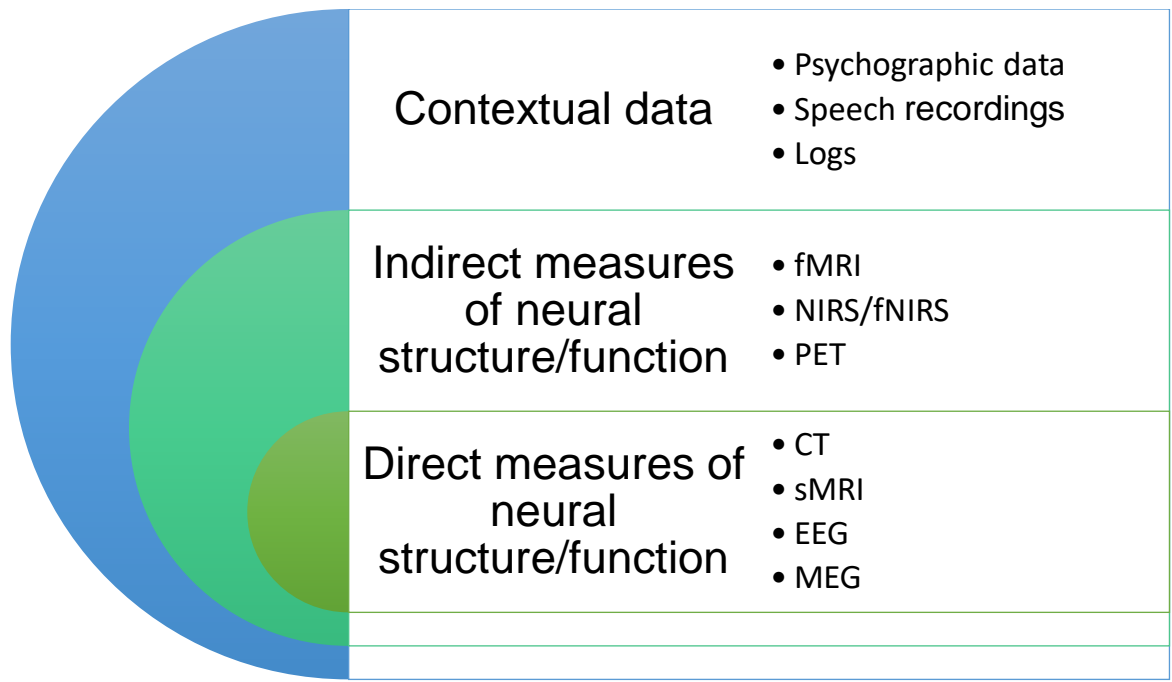

*Figure 1- Brain data taxonomy*

*CT, computed tomography; MEG, magnetoelectroencephalography; EEG, electroencephalography; PET, positron emission tomography; (f)NIRS, (functional)near-infrared spectroscopy; (f)MRI, functional magnetic resonance imaging, (s)MRI structural MRI. The first category consists of methods for directly measuring electrical activity associated with neuronal activity. The second consists of methods for indirectly measuring neuronal activity, which operate under the principle that neural activity is supported by increased local blood flow and metabolic activity. The third class consists of active or passive digital phenotyping data related to perception, cognition, emotion and behavior. The data types presented in this taxonomy should be considered as explicative of each data category, not as an exhaustive typology.*

At the neurobiological level, brain data are the most direct correlates of mental states. Although current neurotechnologies, especially non-invasive techniques, are not yet able to *decode thoughts* —in the sense of providing a full, granular and real-time account of the neural patterns of specific cognitive processes—they allow to infer the engagement of some perceptual and cognitive processes from patterns of brain activation, a process known as *reverse inference* [17]. This occurs through invasive and non-invasive methods to record (and manipulate) neuronal circuits as well as ML-driven data analytics. In laboratory animals, it is now possible to decode visual perception and manipulate it with high precision [18, 19]. In studies with human subjects, researchers have used fMRI scans and high-density electrocorticography signals to accurately decode mental imagery and silent speech [20, 21]. Recent work on intracranial EEG recordings of speech-related brain activity has achieved remarkable accuracy in identifying brain activity patterns related to inner speech [22] while ML techniques have helped enhance the analysis of cognitive processes also from EEG measurements [23, 24].

Finally, research has shown that predictive inferences about mental states can be drawn also from non-neural data sources such as behavioural and digital phenotyping data [25]. Since network neuroscience models and ML techniques are increasingly acquiring inferential power, brain data analytics will likely result, in the long term, in a greater disclosure of mental information. Big data approaches combining brain data and contextual data may offer additional inferential resources for such predictive analytics and allow for more far-reaching and personalized inferences, especially regarding mental content. Mental decoding can improve our scientific understanding of mental illness and holds promise for the targeted modulation of mental states. At the same time, it raises privacy and security challenges.

Even without decoding mental information, current inferential models based on brain data can make privacy-sensitive inferences about present and future brain function or health status. These inferences and predictions, including early signatures of cognitive decline, can be made about both individuals and groups [26]. Since brain data can be stored digitally, more information will become inferable in the future, as scientific understanding of brain processes and decoding algorithms improve. Furthermore, brain data have higher temporal resolution and potential for real-time interaction compared to other biomedical data such as genetic data. This enables more time-sensitive access to brain activity, e.g. for real-time brain-computer interfacing. Finally, brain data are not "read-only" but are often available in a "read-and-write" format due to neuromodulation such as via electromagnetic brain stimulation techniques and optogenetics. This opens the prospect of targeted and direct influence on a person's mental life and personal identity.

It should be highlighted that many neurotechnologies currently available in the consumer space have limited precision [27]. However, with the current pace of technological progress, increasing market growth and the frequent spillover of biomedical technology into the non-medical sector, brain data processing for non-medical purposes raises the need for anticipatory ethics and foresight governance.

**Ethical and legal challenges of brain data**

These unique properties of brain data raise substantive ethical and legal challenges. Since the human brain governs not only life-maintaining physiological processes but also cognitive, affective, volitional, and social faculties [28-30], brain data raise challenges for fundamental normative and legal constructs such as personal identity, autonomy, freedom of thought, moral agency, mental privacy and mental integrity. The notion of "freedom of thought", for example, has been historically characterized as the right and freedom to protect the externalizations of thought such as choice (*freedom of choice*), language (*freedom of speech*) and behavior (e.g. *freedom of expression*). Brain data processing may solicit a literal reinterpretation of the right to freedom of thought. Similarly, the notions of personhood and personal identity are highly dependent on individual brain function and directly affected by changes to brain activity via neuromodulation.

Further, brain data processing raises novel challenges for the notion of mental privacy for two reasons. First, privacy is predicated upon the conscious ability of the individual to filter the flow of data and intentionally seclude private information. Brain data, in contrast, are mostly elusive to conscious control, hence cannot always be intentionally secluded. While this problem is shared with other data types (e.g., genetic data), it acquires greater ethical complexity in the neural domain. Specifically, brain data admit no separation between the data processed and the system that makes decisions about their processing (the human brain). Second, brain information is the ultimate resort of informational privacy since it includes unexecuted behavior, inner speech or other non-externalized action. In principle, mental privacy can be preserved even if individual behavior is constantly surveilled through activity tracking, personal digital technology, self-quantification or simple observation. It could be argued that when one agrees to allow brain data to be acquired, one seems to surrender the right to mental privacy, at least to some degree. However, in scenarios where brain data collection is either mandated (e.g. in the military sector or workplace) or competitively advantageous (e.g. Facebook's plan to make brain-typing faster than the touch-screen), the risk of sharing data under explicit or implicit coercion is concrete.

AI-driven brain data processing may allow access to mental information and bring privacy debates into partially uncharted territory. Legal systems are well-equipped to protect the '*locus externus*' (behavior, verbal utterances, written text etc.) but less-equipped to protect the '*locus internus*' (e.g. unspoken information, preconscious preferences, attitudes, and beliefs). Data subjects may lose control over their brain data in several ways: (i) by consenting to the collection of their data without being adequately informed (e.g. on a device's Terms of Use due to the complexity of the subject matter); (ii) by providing informed consent to the processing of their data for a certain purpose but remaining unaware of further reuses of their data for different purposes (including scraping by third parties); (iii) by being coerced to have their data collected (e.g. via employer's mandate or in an interrogation context).

The nature of brain data might also compromise the ability of data subjects to exercise their rights to access, edit and delete their own data. For example, a data subject might not possess a computer powerful enough to process data from a BCI [31]. Likewise, deleting brain data may substantially decrease the accuracy of ML models generated with these data. Finally, brain data processing generates a risk of "neurodiscrimination", i.e., discrimination based on a person's neural signatures (indicating, for example, a dementia predisposition), or mental health, personality traits, cognitive performance, intentions and emotional states.

**Gaps in the current ethical, legal framework**

We identify four intimately interconnected areas that require attention and proactive governance to ensure the safe and responsible use of brain data outside of the biomedical domain:

**Gaps in supranational and international law:** No mandatory governance framework focused on brain data currently exists in supranational or international law. *Prima facie*, brain data are personal data, as codified inter alia in the legally binding European Union's *General Data Protection Regulation* (GDPR), the non-binding 2013 OECD's *Privacy Guidelines* and the upcoming Council of Europe's (CoE) *Modernized Convention for the Protection of Individuals with Regard to the Processing of Personal Data*. Under these instruments, personal data are defined as any information related to an identified or identifiable natural person (Art. 4 GDPR; Art. 1 OECD Privacy Guidelines, Art. 2a CoE).

However, there are a number of limits with this definition in the context of brain data. Firstly, the GDPR is not applicable if brain data are anonymized even though the technical difficulty of anonymizing brain data leaves open the potential for re-identification. Research shows the feasibility of re-identifying data subjects based on electrophysiological measurements or neuroimaging data, predicting present emotional states and future behavior from brain data, as well as decoding information either from the neural activity of data subjects or their digital phenotypes [24, 32]. Because of the technology involved in the processing of brain data and its high contextualization, the likeliness that anonymized brain data (or data *thought to be anonymized*) will become re-identifiable is non-negligible.

Secondly, unique characteristics of brain data pose challenges to safeguarding the rights of data subjects. A prominent example is the *right to be forgotten*, i.e. one's right to request a data controller to delete his/her personal data. A key characteristic of brain data is that they are potentially re-identifiable and elude conscious control. Therefore, even if a person is initially able to have their data deleted, the data controller or others might use those data to derivatively reconnect them to the person concerned. Most importantly, in the case of brain data involving unconscious information, the data controller might be able to retain data the individual is not aware of. Finally, data deletion by consumer BCI companies may be difficult to obtain due the impact that such erasure would have on the accuracy of predictive models [31].

Thirdly, the GDPR allows some derogations to the rights of data subjects when special data categories are processed for research or statistical purposes, even when the research is conducted by a private company. This implies that some processing of brain data by both public and private actors (e.g., government agencies or consumer neurotechnology companies), may rely on derogations from the main GDPR rules. In fact, it is unclear under which conditions the privilege to use brain data for scientific research (as defined in Article 5 (1) (b) GDPR) applies to brain data collected in the consumer context. Increased transparency is therefore necessary to respect various research purposes and to allow data subjects to intervene into the further processing if this occurs for undesirable purposes.

Further, brain data may undermine another principle of data protection law, namely *purpose limitation*. By default, sensitive categories of personal data (including health data) can only be collected for specific purposes that need to be specified at the time when consent is given by the data subject. However, purpose limitations are very difficult for brain data because current technology cannot preemptively discern purpose-specific data from the myriads of brain signals. Data security measures that intend to balance consent to broad processing purposes are difficult to define to the law threshold of re-identifiability. Tools for selective filtering such as the *Brain-Computer Interface Anonymizer* are in early stages of development[33].

Finally, safeguards provided by data protection law may not adequately scale to group-level data. This lack of adequate scaling raises a twofold group-privacy risk: first, third parties can make inferences about a group of data subjects based on one or multiple features inherent in the brain data and shared by all individuals in the group (e.g. slower reaction time to cognitive tests). Second, individuals could be unwittingly identified through their brain data, however anonymized, as part of a hitherto unsuspected group (e.g. people showing prodromal signatures of cognitive decline) and subsequently discriminated against.

To complicate things, brain data generated from consumer neurotechnologies may not constitute 'health data' hence are subject to lower protections compared to data from clinical applications because the application of these devices does not fall under medical device regulation regimes [34].

**Gaps in ethics and soft law:** The collection and processing of brain data within biomedical or clinical research is further governed by research ethics guidelines for the protection of human subjects. These include the *Belmont Report* and the *Declaration of Helsinki* by the World Medical Association, as well as through oversight mechanisms such as Institutional Review Boards. These instruments are critical to uphold the rights and responsibilities of the research community in the conduct of biomedical and clinical research. However, they do not apply in the consumer, neuromarketing, workplace or military domains. In the consumer space, simply prompting users to accept a service's Terms of Use places the responsibility on users to understand these terms and does not guarantee informed decision-making[35]. Even if consent can be obtained in a broad manner, current ethical safeguards are ill-suited to guide the entire data lifecycle. This is particularly true given the trend towards perpetual recycling and re-contextualization of previously collected data [36]. Further, ML allows to draw post-hoc private and confidential inferences from non-sensitive data, prompting further need for the protection of data subjects [12]. Based on these considerations, experts have called for ethical guidelines for novel consumer neurotechnologies to fill persisting gaps in data governance [37].

**Gaps in responsible innovation:** Currently, most applications that collect and process brain data outside the clinical and medical research context do not seek compliance with the EU Medical Device Regulation (2017/745) or approval from the US Food & Drug Administration (FDA). Approval from these agencies is only necessary for software and devices with a medical purpose. This bypassing of the relevant medical device regulation is generally predicated on the non-medical scope of these devices and programs. However, consumer and military neurotechnologies can collect medically relevant parameters (e.g. via EEG measurements) and often claim to draw inferences about cognition or psychological wellbeing. Many wearable devices and applications are available for commercial, personal and even health-related use without relevant labelling required by data quality standards [27]. Typically, users of consumer neurotechnology devices or services have no information about how in-house brain function databases are compiled. Further, users have no guarantee that such databases are sufficiently representative to provide valid assessments of individual or group-level cognitive function and affective state [38]. Insufficiently validated applications may incorporate bias, provide false information or even cause harm to the users such as when users make health-related decisions based on these apps. Additional hazard may be posed by malicious hacking, eavesdropping, unauthorized access by third parties, unsecured data transmissions, re-identification of anonymized data and identity theft. Some of these risks also extend to the clinical and biomedical research field. An important step towards innovation governance was recently marked by the Recommendation on Responsible Innovation in Neurotechnology, which was adopted by the OECD in December 2019, setting the first international standards for responsible innovation in this domain [39].

**Gaps in international human rights frameworks and further lacunae:** Human Rights instruments, such as the UN's *Universal Declaration of Human Rights* (1948), which is legally binding as part of customary international law, were drafted long before brain data became measurable outside the clinic and amenable to big data analytics. Given this, they did not explicitly spell out requirements for gaining access to and using brain data in a manner that protects individual rights. Whereas the conditions for legitimate use of human genetic data have been delineated in UNESCO's soft law *International Declaration on Human Genetic Data* (2003), human brain data remain without explicit safeguards and lack comparable protection by human rights instruments. Further, there is no specific international treaty that addresses the dual-use or potential weaponization of brain data for military purposes. Dual-use research and technology collecting human brain data is therefore a pressing anticipatory governance concern as neurotechnology evolves and is increasingly researched in the military setting.

**Towards a Multi-Level Governance Framework**

Advancing the use of brain data in neuroscience and medicine while simultaneously preventing ethical-legal risks requires a delicate balancing act. As brain data intersect several domains of human activity and regulation, it is unlikely that a one-size-fits-all approach to governance can be effective. Therefore, a comprehensive framework for global governance should operate adaptively at multiple levels. Based on the previously identified gaps, we propose four primary areas of regulatory intervention: binding regulation, ethics and soft law, responsible innovation, and human rights (Figure 2).

A. **Binding Regulation:**

Mandatory governance efforts seek to define and locate brain data within the supra-and-international data protection landscape. We suggest that brain data should be considered a special category of personal data that warrants heightened protection during collection and processing. We posit that singling out brain data as a special category would help govern the non-medical use of these data while safeguarding their processing for scientific and biomedical purposes. This approach is consistent with the risk-based approach of the GDPR and could mimic the framing of other special categories of personal data such as genetic data (which includes chromosomal, DNA or RNA data; Article 4(13)). This would allow to protect brain data also prior to analysis, when they cannot be linked back to an identifiable individual or when they are generated by non-medical devices.

Additional provisions may clarify conditions for collecting and processing brain data in the non-medical space. At the data privacy level (e.g. as according to the GDPR), device and software manufacturers should ensure data protection "by design and by default" (GDPR, Article 25). Further, data processors and controllers should use pseudonymization and encryption to guarantee data security (GDPR Articles 32-34) and implement the principle of data minimization. Additional measures may include protecting against third-party apps linked to consumer neurotechnology applications. Finally, the exact conditions and safeguards under which the research exemptions, introduced by Union or Member State law on the basis of Art 89 (2) GDPR, can permit brain data processing by private companies should be clarified. To fill a gap in international regulation, we contend that brain data indicating neurological or mental illness originating from non-medical neurotechnology should not be accessible by third-party actors such as health insurance providers. Access to such information would require the user's explicit and written (or digitally provided) consent.

More broadly, risks for privacy and human dignity specific to brain data analytics must be disclosed. In particular, regulators must consider whether a right to *mental privacy* and *mental integrity* should be granted to data subjects. These rights would grant subjects increased control and protection of data containing information about their sensory, cognitive, affective and volitional processes. In addition to data protection law, criminal and civil laws could reinforce these privacy rights by protecting a person's brain activity against unconsented exploration and modulation. Labor law offers grounds to protect employees from the misuse of their brain data in an employment context, e.g. by prohibiting employers from collecting brain data for productivity monitoring and terminating employment contracts based on brain data.

Another critical issue is the coercive collection of brain data. Governance frameworks should protect the ability of people to make free and competent decisions about the collection and processing of their personal brain data, a principle known as *cognitive liberty*. The *European Convention on Human Rights* (ECHR), which protects the rights to privacy and freedom of thought (Arts. 8 and 9) offers the suitable conceptual and normative framework to prevent coercive uses. If the CoE Modernised Convention comes into force, it could serve as a solid basis for further specification and a model for other world regions.

In order to increase compliance and promote sustained scientific validation of new devices and algorithms in the gray zone between the medical and the non-medical domain, calibrated amendments to current medical device regulations should be considered. Currently, most consumer neurotechnology companies avoid classification of their products as medical devices by marketing them for wellness, relaxation and other non-medical purposes [27]. Nonetheless, users (including vulnerable people) may use those devices and share their brain data for health-related purposes, such as cognitive monitoring and mental wellbeing. A step towards reform was taken by the EU's amendments to the *Medical Devices Regulation*. These amendments will apply from May 2021 and cover also brain stimulation products without an intended medical purpose as medical devices (Annex XVI, No 6). However, it does not cover brain data processing for purposes other than neuromodulation. Furthermore, it remains highly uncertain whether and how regulatory agencies will take enforcement action.

Apart from peaceful purposes, the limits of exploring and modulating brain function for military usages must be defined. This is especially relevant as large military research agencies, such as the DARPA in the US, actively pursue brain stimulation technologies for modulating cognitive functions, such as memory and learning [5]. In an international context, brain data (as the decisive parameter for calibrating such neuromodulation devices) could thus become a commodity in a neurotechnology "arms race" as other nations also pursue military neurotechnology research and development. This arms race could involve both the development of novel military neurotechnology and the dual-use (repurposing) of consumer or medical technology [6]. The laws of war that are applicable during armed conflict [40] (so-called *international humanitarian law*) do not explicitly protect combatants against the violation of their mental integrity. There is a need to draft legal guidelines—similar to those guiding autonomous weapons—that protect soldiers against brain data misuse during both wartime and peacetime.

**B. Ethical Guidelines and Soft Law**

Despite the difficulties of cross-border data transfers, brain data sharing practices are generally not restricted by national borders and regulatory frameworks. Therefore, internationally applicable ethical principles and rules are needed to govern the collection and processing of brain data. Research ethics procedures such as review through ethics committees and Institutional Review Boards (IRBs, which, in some countries and some areas of research, are part of binding law) are well-established governance mechanisms for the clinic and human neuroscience research. However, these procedures are insufficiently agile to respond to the novel challenges posed by the current big-data digital ecosystem, especially the innovation dynamics and business models of AI-based technology in the neuroelectronics marketplace. Similarly, the evaluative criteria of ethics review are not geared towards the current information-intensive ecosystem.

We posit that legitimate interest alone is insufficient to provide the ethical basis for brain data processing. In addition, consent should also be considered as critical ethical requirement for a brain data governance framework. This is consistent with the opinion of the European Group on Ethics in Science and New Technologies, which proposed to include individual consent as a requirement for further processing of health data in the EU regulation [41].

When collecting and/or processing identifiable brain data, private data collectors must conduct a legitimate interest assessment, check that the processing is necessary and there is no less intrusive (non-neural) way to achieve the same result. Further, they must document that *explicit* informed consent for a specific usage was obtained prior to data collection except in cases of medical emergency. Data collectors should be required to apply explicit informed consent procedures that go beyond the mere acceptance of ToU for consumer products. These procedures should transparently disclose and address, not less than: (i) how brain data are used, i.e. which information is decoded and with which accuracy; (ii) in which storage facility and on what medium data will be stored and the duration of storage; iii) the criteria and mechanisms by which access to the brain data is granted, monitored and revoked; (iv) how brain data are reused and shared; (v) what anonymization/pseudonymization and information security measures are implemented; (vi) how individuals will be informed if their data are hacked, leaked or accidentally disclosed, and; (vii) what legal entity is liable for data breaches and other regulatory lapses. Novel digital technologies for informed consent (eConsent) have shown potential to enhance the practicability and efficacy of consent procedures [42]. In practice, in a clinical setting or in the consumer space, adherence to these procedures could be governed and monitored by Data Use and Access Committees, e.g. adjunct to IRBs in the clinic or consumer protection agencies.

We argue that the default consent for governing brain data use should be an opt-in approach. Accordingly, individuals have to explicitly *opt in* to sharing their brain data or link this data with other contextual information (e.g. social media profiles). Ethical guidelines should extend beyond mere rule-compliance and promote the respectful use of brain data.

## C. Responsible Innovation

Responsible Research and Innovation (RRI) is now a widely accepted approach for guiding emerging sciences and technologies and promotes first and foremost the responsible collection and processing of brain data by both public and private actors. RRI principles can help develop safer and more reliable systems as well as increase preparedness to deal with unintended consequences. These include the adoption of community-agreed technical standards (e.g., within the neuroengineering community [43]), adequate validation and best practices by neurotechnology researchers, companies and other stakeholders in a consensus process.

Service providers who collect and process brain data should ensure safety, scientific validity, accountability and transparency. At the safety level, usage of brain data should consider and prevent inherent risks of algorithmic processing including bias, privacy violation, and cybersecurity vulnerabilities. Data collectors and processors should ensure data minimization, for instance by only providing data from some EEG channels or by selectively filtering certain frequencies in the data.

Novel privacy-preserving technologies can help both medical and non-medical processors. Technical approaches to improve protection from leakage and unwarranted access include homomorphic encryption, multi-party computation, federated learning, and differential privacy [44]. Differential privacy is particularly well-suited for brain data because it allows sharing aggregate data whilst preventing inferences from being drawn about individuals. Nonetheless, some risks can only be discovered once the systems have been deployed.

Accordingly, developers shall establish mechanisms for continuous analysis, monitoring and mitigation of risk once software and devices are on the market.

Finally, data collectors and processors should ensure high standards of scientific validity for both devices and datasets. Consumer service providers should be prevented from advertising unsubstantiated paramedical claims (e.g. "improving mental wellbeing") that are loosely founded, if at all, on scientific evidence [27]. Adequate testing and careful risk-benefit analysis should guide development and deployment of brain data processing systems. This will likely improve not only the safety, but also the efficacy, user-friendliness and precision of future devices. Similarly, regulators should take a proactive stance on the ethical, legal and social implications of these technologies. This proactive stance requires constant interaction between all stakeholders to identify suitable means for standardization, such as value-sensitive design. Oversight mechanisms involving binding regulation, soft law and ethical guidelines shall make sure that these standards are met by laying down the necessary permit procedures.

### D. Human Rights

Brain data are inherent to and in principle accessible from all human beings, regardless of ethnicity, gender, nationality or religion. Further, they reflect the inner workings of our minds as they correlate with thoughts, emotions and other mental faculties. Therefore, the prospect of unsupervised deciphering of, interfering with and commodifying brain data raises serious human rights challenges. We posit that non-medical brain data processing for legitimate interest should not be pursued when the controller's interest conflicts with the fundamental rights and freedoms of the data subject. Human rights inform legislation, ethical guidelines and societal norms across the globe, and thus offer an international normative framework where brain data protection needs to be embedded. Interdisciplinary research investigating the intersection between brain data and human rights is ongoing and proposals for protecting neuro-specific rights, called *neurorights*, have been advanced [45,44]. Governance frameworks should determine whether clauses can be added to human rights treaties or whether a new universal soft law instrument is necessary. This instrument could be modelled after the 2005 *UNESCO Declaration on Bioethics and Human Rights*. Furthermore, it should be determined whether neurorights should be interpreted as new rights or as adaptive interpretations of existing legally binding human rights and moral principles. These rights include the right to privacy, the right to freedom of thought, mental integrity and human dignity. The Universal

Declaration of Human Rights, in particular, grounds human rights in the inherent and equal dignity of all human beings.

The normative force and universal claim of human rights often makes it difficult to translate these rights into guidance for context-sensitive action. For this reason, frameworks such as the capabilities approach[46] can be very helpful to translate the general requirements of human rights into actionable and shared international policy goals that promote human flourishing, human dignity and well-being in the context of brain data processing. Several national (e.g., Chile's recently approved Constitutional Reform and Neuroprotection Bill) and international organizations actors (the Council of Europe, the EU Parliament, and the OECD) are putting "neurorights" on their agenda.

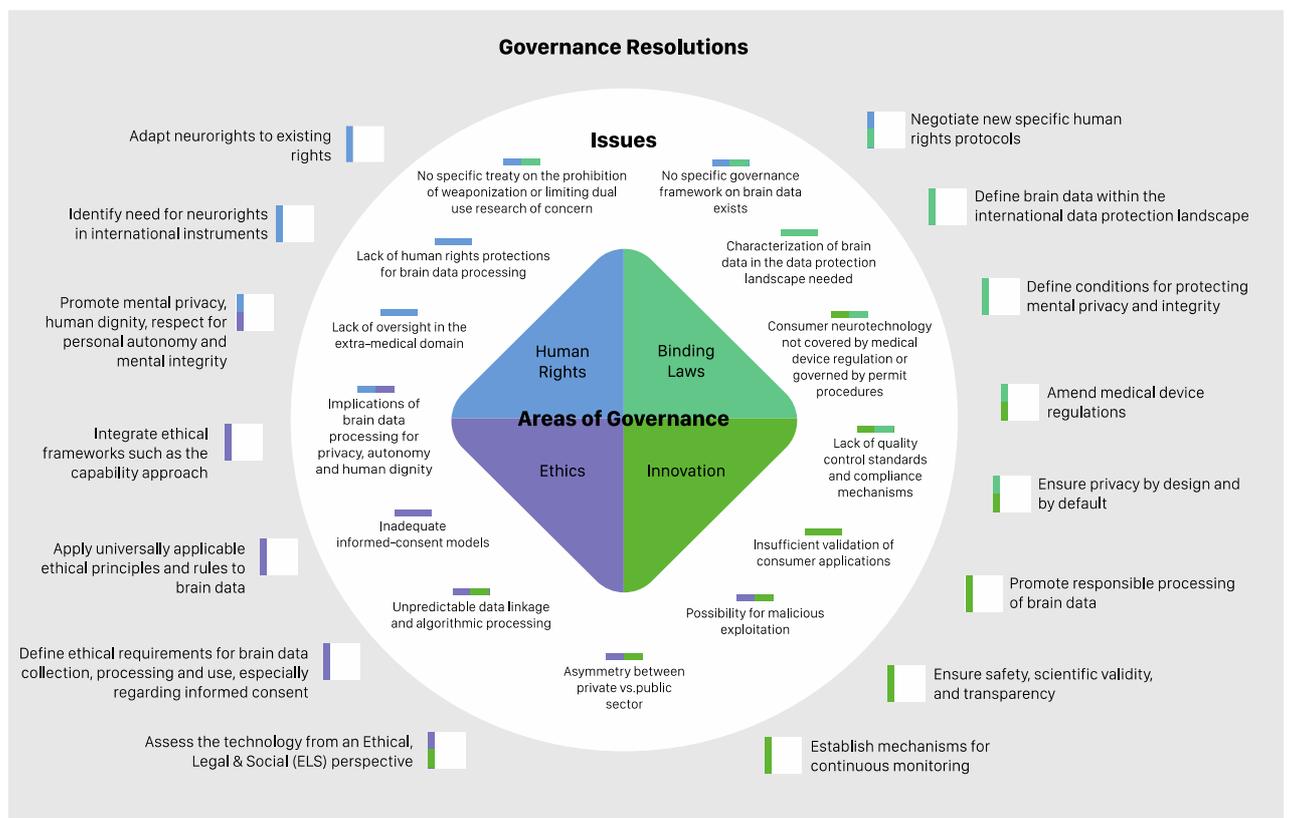

*Figure 2- Overview of normative requirements and levels of governance*

## Conclusions

International governance should ensure the positive impact of brain data processing on science, health, well-being, human dignity and human rights, while preventing potential risks for individuals and communities. We delineate a roadmap towards a global governance framework on brain data that can fill current ethical and legal gaps. We call upon professional societies, national and international organizations, as well as unrepresented or underrepresented communities and stakeholders (e.g. patient organizations) to take up the challenge and coordinate a joint effort at their adoption. Any move towards an international framework should be aware of cultural diversity and responsive to a pluralistic global society. Finally, following recent challenges in AI governance, we should avoid the uncoordinated proliferation of normative guidance in the absence of adequate strategies for harmonization, standardization and implementation.

**Acknowledgments**

*This consensus paper was developed within the scope of a workshop funded by the Brocher Foundation which took place in Hermance, Switzerland, on November 25-27, 2019. The consensus development process involved a series of group activities including interactive panels, expert groups, and plenary sessions.*